# Inverse Design of Photonic Crystal Devices

Andreas Håkansson, José Sánchez-Dehesa and Lorenzo Sanchis

*Abstract*— This work deals with the inverse design in the field of photonic crystal based devices. Here an inverse method containing a fast and accurate simulation method integrated with a competent optimization method is presented. Two designs yielded from this conjunction of multiple scattering theory with a genetic algorithm is analyzed. The potential of this approach is illustrated by designing a lens that has a very low F-number (F=0.47) and a conversion ratio of 11:1. We have also designed a coupler device that introduces the light from an optical fiber into a PC based wave-guide with a predicted coupling efficiency that exceeds 87%.

*Index Terms*— Inverse design, photonic crystals, photonic devices, genetic algorithm, spot-size converter, multiple scattering theory, optical fiber, photonic crystal wave-guide

## I. INTRODUCTION

PHOTONIC CRYSTALS (PC's) [1] are promising materials in order to produce new devices. Different compact optical devices and circuits can be designed by introducing point and/or line defects in these crystal structures. Although the predicted performance of these photonic devices is very promising, many practical problems still need to be solved before they will be competitive in the market-place. To solve these problems a normal approach is to use trial and error guided by knowledge and intuition. Even if this direct design technique is useful and relatively effective it is strongly restricted to the degree of insight into the problem. This is one great limitation using direct design. Through direct design it is impossible to invent a device without being able to give a complete description of the underlying physics of its functionality. To find unexpected non-trivial solution that might be overseen in direct design, the trial and error should be controlled, not by intuition, but by a stochastic or deterministic optimization algorithm. The physical intuition is then used to limit the search space and lower the diversity of possible solutions by setting up constrains.

To apply an optimization algorithm it is convenient to be able to simulate the functionality of the device we are designing. The inverse design (ID) problem is solved by optimizing the parameters within the set constrains. Solving problems using inversion methods has been demonstrated to be very effective for various PC-devices such as (other than the two analyzed here); PC's for cavity QED experiments [2], PC's with large absolute band gaps [3], aperiodic scatterers for tailored transmission or reflection [4], and low loss PC wave-guide Z-bend [5].

A. Håkansson, J. Sánchez-Dehesa and L. Sanchis are with the Wave-phenomena Group within the Nano-Photonics Technology Center and Dept. of Electronic Engineering, Polytechnic Univ. of Valencia, C/ Camino de Vera s/n, E-46022 Valencia, Spain.

## II. INVERSE DESIGN METHOD

When setting up a problem for ID resolution the most important thing is to enclose the search space in an adequate manner. On the one hand, when setting up the boundaries it is necessary to exclude solutions that we are not interested in, e.g. solutions that we later lack the possibilities to fabricate. On the other hand, it is equally important not to exclude solutions that might lead to an optimal result. Light constrains is preferable since this will include a larger number of non-trivial solutions. The light constrains will lead to a great number of possible designs. This imply a large number of numerical simulations during the optimization process, which require fast calculation methods. Another important matter is the estimation of the functionallity of the device. During the simulation, it is necessary to be able to evaluate this prerequisit comportment. This evaluation is set equal to the parameter of optimization. If this parameter is badly chosen a lot of efforts will be lost in the search looking for non-optimized devices. The method presented here has been introduced by some of us [6], and can be applied to any two dimensional (2D) PC ID problem containing clusters of dielectric or metallic cylinders and works for both TE and TM polarized light. The method is more effective for solving problems with a finite number of solutions, using bounded discrete parameters. Within this wide field of applications, this work concentrates on PCs of silicon cylinders ($n_{Si} \sim 3.46$ at $\lambda = 1.5\mu m$) in a background of silica ($n_{SiO_2} \sim 1.45$) and TM polarized incident light (in-plane magnetic field). Both the simulation and the optimization method are well known in the literature. A description and motivation of choice as well as limitations is given in the two following subsections.

### A. Simulation Method

Here we look at 2D systems containing arrays of dielectric cylinders with circular cross section. The multiple scattering theory (MST) is both a fast and accurate method when simulating this type of crystals. The method has been successfully applied in the analysis of metallic and dielectric clusters based on 2D PCs [7], [8]. The simulation time is dependent on the number of cylinders in the simulation. If considering systems that do not include a to great number of scatterers ($< \sim 200$) the MST can be considered very fast (a system of 100 scatterers takes $\sim 2$ second to simulate on a Pentium IV computer).

In brief, the method uses the T-matrix [9] separately defined for each scatterer to calculate the total scattered wave from the system. To further illustrate, let us consider a cluster of N scatterers located at the positions $\vec{R}_\beta$ ($\beta = 1, 2, ..., N$) = $\begin{pmatrix} x_\beta \\ y_\beta \end{pmatrix}$. If an external wave $E^{ext}$ with temporal dependence $e^{-i\omega t}$ impinges the cluster, the total field around cylinder $\alpha$ is a superposition of the external field and the field scattered by the rest of the cylinders in the cluster:

$$E_\alpha(x,y) = E^{ext}(x,y) + \sum_{\beta \neq \alpha}^{N} E_\beta^{scatt}(x,y), \quad (1)$$

where $E_\beta^{scatt}$ is the wave scattered by cylinder $\beta$. These three fields can be expanded into series of orthogonal Bessel functions. Using the solutions to the scalar wave equation in polar coordinates as base functions, the total incident wave to the $\alpha$ cylinder and the external incident plane wave can be expressed in terms of the Bessel function of the first kind. The scattered wave from cylinder $\beta$ is expressed in a series of outgoing Hankel functions rather than regular wave functions. Using the multi pole coefficients $(B_\alpha)_l$, $(S_\alpha)_l$ and $(A_\beta)_l$ for $E_\alpha$, $E_\alpha^{ext}$, and $E_\beta^{scatt}$, respectively, the expression above can be cast into the following relation between coefficients:

$$(B_\alpha)_l = (S_\alpha)_l + \sum_{\beta=1}^{N} \sum_{l'=-\infty}^{l'=\infty} (G_{\beta\alpha})_{ll'} (A_\beta)_{l'} \quad (2)$$

$G_{\beta\alpha}$ being the propagator from cylinder $\beta$ to $\alpha$ and whose components are

$$(G_{\beta\alpha})_{ll'} = (1 - \delta_{\alpha\beta}) e^{i(l'-l)\theta_{\beta\alpha}} H_{l'-l}^{(1)}(\kappa_1 r_{\beta\alpha}) \quad (3)$$

where

$$\theta_{\beta\alpha} = \arctan(\frac{x_\alpha - x_\beta}{y_\alpha - y_\beta})$$
$$r_{\beta\alpha} = \sqrt{(x_\alpha - x_\beta)^2 + (y_\alpha - y_\beta)^2}$$
$$\kappa_1 = \kappa_0 \epsilon_1 = \frac{\omega}{c_0} \epsilon_1$$

and $\delta_{\alpha\beta}$ is the Kronecker delta, $\epsilon_1$ is the dielectric constant of the medium of propagation ($\epsilon_1 = 1$ for air), $c_0 = 3 \times 10^8 m/s$ and $H^{(1)}$ is the Hankel function of the first kind.

Notice that the coefficients $S_\alpha$ are known, but $B_\alpha$ and $A_\beta$ are not. Using the boundary condition at the interface of each scatterer we can relate $B_\alpha$ with $A_\beta$. This relation, expressed in matrix notation as $B_\alpha t_\alpha = A_\beta$, is the T-matrix and takes the following values for circular scatterers:

$$(t_\alpha)_{ll'} = \frac{\kappa_1 J_l'(\kappa_1 R_\alpha) J_l(\kappa_2 R_\alpha) - \kappa_2 J_l(\kappa_1 R_\alpha) J_l'(\kappa_2 R_\alpha)}{\kappa_2 J_l(\kappa_1 R_\alpha) H_l'(\kappa_2 R_\alpha) - \kappa_1 H_l(\kappa_2 R_\alpha) J_l'(\kappa_1 R_\alpha)} \quad (4)$$

where $R_\alpha$ is the radius of cylinder $\alpha$, $\kappa_2$ and $\kappa_1$ are the wavenumbers inside ($\kappa_0 \epsilon_2$) and outside ($\kappa_0 \epsilon_1$) the cylinders, respectively, and the prime indicates the derivate Bessel function defined as; $\Upsilon_l' = \Upsilon_{l-1} - \Upsilon_{l+1}$, where $\Upsilon$ is either $J$ (Bessel) or $H$ (Hankel).

Now we can print out the system of equations by introducing the coefficients from Eq. 4 in Eq. 2

$$\sum_{\beta=1}^{N} \sum_{l'=-\infty}^{l'=\infty} (A_\beta)_{l'} \left( \delta_{\alpha\beta} \delta_{ll'} - (t_\alpha G_{\beta\alpha})_{ll'} \right) = (t_\alpha S_\alpha)_l \quad (5)$$

By truncating the angular momentum within $|l'| \leq l_{\max}$, Eq. 5 reduces to a linear set of equations where the dimension of the relevant matrix is $N(2l_{\max} + 1) \times N(2l_{\max} + 1)$. It is important to stress that the size of this matrix increases with the number of scatterers, which will result in a notably increase in the simulation time.

The theory gives an exact solution to the system and the result is expressed in $N$ infinite series expansions of Hankel functions ($N$ point sources). Because of the circular properties of these functions the terms in the series expansion converge relative fast for cylindrical scatterers. For example, for frequencies in the first photonic band, the highest significant index is of the order 3. This fast convergence contributes to a very fast simulation time and low CPU costs. The MST apply very well for systems with freely placed cylindrical scatterers, including systems with cylinders with different radius. Another convenience is that if the system and the incident wave is symmetric with respect to one axis the T-matrix of total system can be cut to a fourth [10] resulting in a considerable simulation speed-up.

It has to be mentioned that this simulation method might not be the proper choice for arbitrary shaped scatterers. If one lifts this constrain and gives more freedom to the shape of the scatter the T-matrix used in MST must be recalculated for every different formed scatterer. In this regard the finite element method has been proved useful [5].

### B. Optimization Method

The No-Free Lunch theorem by Wolpert and Macready [11] states that for any algorithm, any elevated performance over one class of problems is exactly paid for in bad performance over another class. In other words, it does not exist no better nor worse optimization algorithm with respect to its average performance on all possible classes of problems. Fortunately optimization problems connected to real applications almost always belong to a class of functions with elevated performance. The verity in optimization algorithms is large and can generally be divided into two groups, deterministic or stochastic algorithms. Deterministic optimization is characterized by the fact that there always exist at least one instruction associated with a given present state of the search, while stochastic search algorithms use a pool of solutions and guides its way using comparison. Determinisic optimization is often used when dealing with nice and smooth optimization problems of less complex nature, while stochastic search is more frequently used to solve harder combinatorial problems. Further, stochastic search show great flexibility and robustness and can easily be integrated in mixed and hybrid search strategies. If stochastic search is more effective for finding maximas it generally uses a large number of function evaluations to do so. This inconvenience favor fast simulation method such as the MST presented in the previous section.

One popular family of stochastic search algorithms, frequently applied to solve engineering problems, is evolutionary computation. This assemblage of algorithms are related with the optimization process used by nature itself, the evolution. The evolution is very powerful in adapting individuals to a given environment, and is able to tackle enormous complex problems with fairly simple means. Throughout generations the individuals' chromosomes (the genotypes) are mixed and new individuals with different characteristics are born. Those individuals that adapt better to the environment have the best chance of survival and hence give birth to more offspring creating a new generation more fit for survival than the previous one. To judge how well individuals adapt they are associated with a fitness value. This fitness reflects the wanted functionallity of the optimized device.

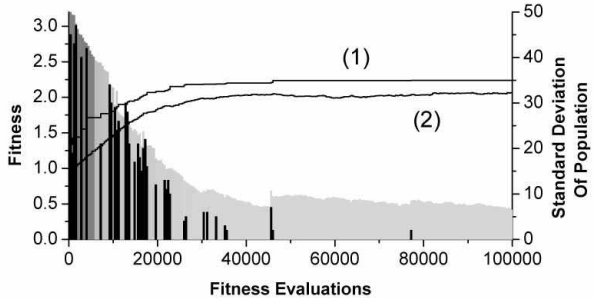

Fig. 1. A typical GA-run optimizing a problem coded with 100 binary parameters. Line (1) and (2) defines the best and average fitness, respectively, of the population at a given generation. The gray colored bars indicate the average number of bits a solution differs from the best fitted individual in the population at a given stage. The black bars indicate how many parameters are changed at each best-fitness increase.

We here use one of the most popular algorithms from this family, a simple-GA, introduced by Holland [12]. The GAs are population based stochastic search algorithms normally used to solve discrete problems [13]. Their functionality is based on their ability to find small parts of the individual (the device) that reflects a good quality in the result, these fractions are called building blocks (BBs). In other words, the GA build, mixes and finally uses these BBs to find a global optimal solution. Unfortunately this goal is not always met. If the global optimum consists of large sized BBs, the GA will have problems of constructing them because high-order BBs are very difficult to build and maintain throughout the search. These problems are referred to as GA-hard problems [14] and can be bypassed by representing the problem differently or by trimming the GA-parameters. However, when such problems need to be tackled, a more dynamic algorithm should be at hand, e.g. cGA (compact GA), fmGA (fast messy GA), BOA (Baayesian Optimization Algorithm) [14] that are capable of dynamically recoding the genotype and in this way finding larger sized BBs.

The simple-GA works with three operators; selection, crossover and mutation that are iteratively applied to a population of individuals. The selection operator culls the population selecting solutions with a high fitness for mating. The mating act, which is directed by the crossover operator, mixes and constructs the BBs from two or more individuals creating new offspring. The mutation operator makes it possible for the offspring to possess new BBs that none of the parents have and is applied before passing into the next generation (i.e., the next iteration). The individual is represented by a chromosome (a digital string) that is put together by a number of binary genes coding the genotype. Each gene corresponds to one specific part of the phenotype, i.e. one gene codes one parameter of the problem. The genotype represents one and only one phenotype that here is a cluster of dielectric cylinders. Figure 1 illustrates a typical GA-optimization. Notice that the convergence is methodically reached after $\sim 50'000$ simulations.

The binary implementation leads to the inconvenience of excluding float parameters in the optimization. One normal approach to handle real numbers is to implement a smallest step approximation. This on the other hand can lead to complications since, e.g., 7 (as a binary representation: 0111) and 8 (1000) are following integers and probably very similar phenotypes but differ all 4 digits on the chromosome (genotype). A useful phenotype-genotype representation in this case are Gray-codes [13]. A second approach is to leave the binary representation and use a real-coded GA.

## III. RESULTS

The applicability of the method described above is illustrated by solving the inverse problem of a photonic lens with flat surfaces and a PC wave-guide coupler. Both structures show new interesting features; The lens device focuses the light into a region that is one order of magnitude smaller than the incident beam; The insertion loss predicted for the coupler is about 13%, which is one of the lowest reported by numerical simulations.

### A. Photonic lens with flat surfaces

Let us deal with the design of a photonic lens with flat surfaces. We are here looking for a device with a large $x$-component of the Poynting vector in the chosen focal point, hence the fitness parameter of the optimization is set directly proportional to this value. Now we need to encapsulate the problem by setting up propper constrains and so limiting the search pool. First, the simulation method favor the use of cylindrical scatterers. Secondly, the binary representation used in GA optimizations induces fixed position and constant radius for the cylinders. Hence the only tunable binary parameter in the problem is set to the presence or absence of each cylinder.

The lattice points corresponding to the cylinders' positions have to be chosen carefully. The number of solution increases as $2^N$, if $N$ is the number of lattice points. An increase in $N$ implies an increase in the optimization time needed to solve the problem. Consequently N has to be chosen as small as possible but at the same time in such a way that no optima are excluded. A Gaussian shaped beam is chosen as an incident beam, representing the light from a fiber. The beam, with a width of $2w_o$, is centered at the origin of the coordinate system and propagating along the positive $x$-axis. Consequently, the size of the crystal lattice along the $y$-direction has to be chosen slightly larger than the incident beam's width to be sure that all the light passes through the device. The lattice has a total width of $3w_0$ along the $y$-axis. Regarding its thickness along the $x$-axis various optimizations were done. Here we present the result for 9 and 13 layers, based on a hexagonal symmetry lattice. For 13 layers the total number of lattice points is 318, which is conditioned by our calculation resources. The polarization is chosen as in-plane magnetic field (TM). This choice of polarization is arbitrary since TE polarization can be implemented in this same way. The chosen hexagonal lattice constitutes the lens-material (LM) with the lattice constant $a_{LM}$ and the radius is set to $r = 0.294 a_{LM}$. A band structure calculation of the corresponding infinite crystal by means of a plane wave expansion method predicts a forbidden gap for the TM-like modes and the $\Gamma K$ direction, in the range of frequencies 0.222-0.292 (in units of $2\pi c/a_{LM}$), see left panel in fig. 4. In order to minimize the reflectance we choose a working frequency of 0.197; i.e. below the first gap, where propagation is allowed in all directions for the corresponding infinite crystal. The CPU time needed to get the fitness of one 13 layers individual is of the order of 6s in a 2.8 G Hz Pentium IV workstation. The coordinates of the focal point $(x_f, y_f)$ are freely determined. Particularly, we impose the focal point to be located on the symmetry axis of the system ($y_f = 0$), and with a $x$-coordinate $x_f = 4.60\lambda_0$, where $\lambda_0$ is fixed by the working frequency ($\lambda_0 = a_{LM}/0.197$).

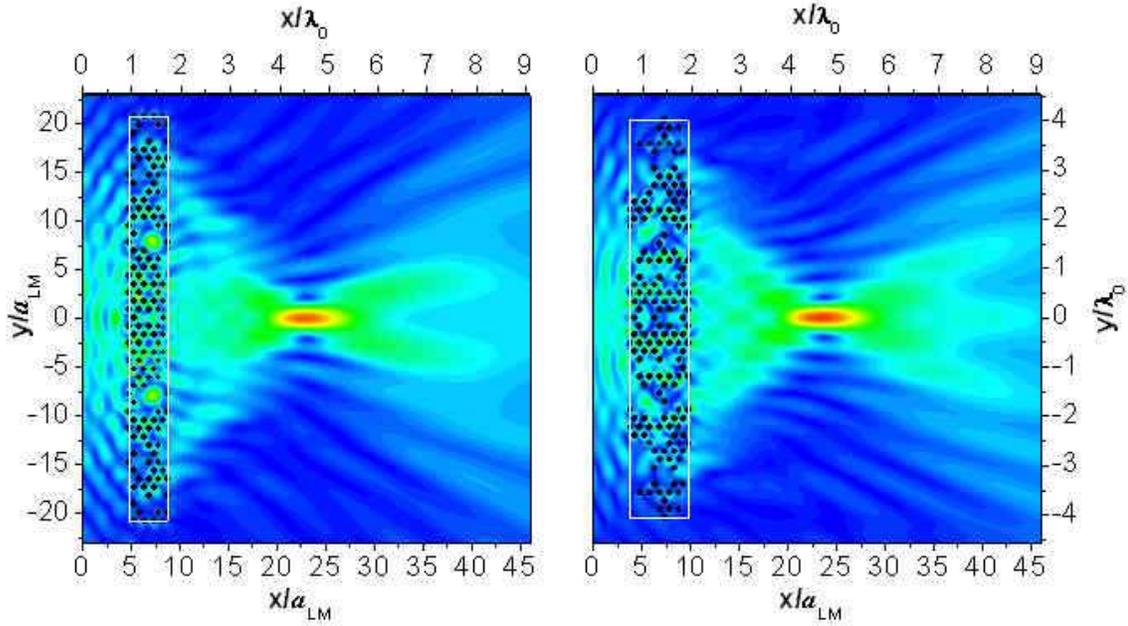

Fig. 2. Focusing effect produced by a lens based on a 2D photonic crystal (black dots). The pattern of the electric field modulus is represented in a wide spatial region. The length scales are given in terms of the lattice parameter employed in the design of the lens, $a_{LM}$, as wells as in terms of the working wavelength of the lens $\lambda_0$. A scale color is used; red (blue) color means maximum (minimum) electric field modulus.(Left panel) The inversed design 9 layers lens structure and (Right panel) the corresponding 13 layers lens.

The global maximum must correspond to a symmetric structure, which is obvious from the symmetry of the problem. By removing all non-symmetric crystal solutions the space search size is reduced from $2^N$ to $2^{(N+n)/2}$ possible configurations, where $N$ is the number of lattice-points, and n the number of lattice points centered on the symmetry axis. This means for the 13 layers lens problem, where $N = 318$ and $n = 6$, a reduction from $5.3 \times 10^{95}$ possible configurations to $5.8 \times 10^{48}$ and for the 9 layers problem from $1.7 \times 10^{66}$ to $2.1 \times 10^{34}$. The left (right) panel in fig. 2 shows the resulting lens device for 13 layers (9 layers) and the pattern of the corresponding electric field modulus. Two different scales are used to measure distances, lattice parameter units (bottom and left) and wavelength units (top and right).

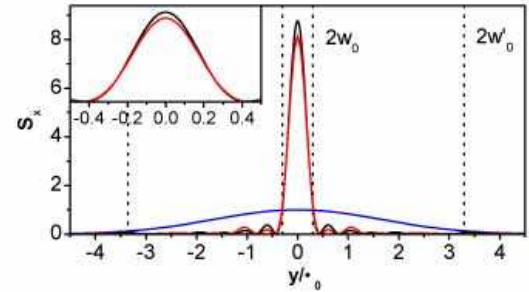

Fig. 3. The $x$-component of the Poynting vector represented at the focal point ($x_f = 4.60\lambda_0$) for the two structure presented in fig. 2. The blue, red and black line line correspond to the width of the incident beam (w'$_0$), the 9 layers lens and the 13 layers (w$_0$) lens respectively. (inset) A zoom over the $x$-axis showing the response from the two lens structures

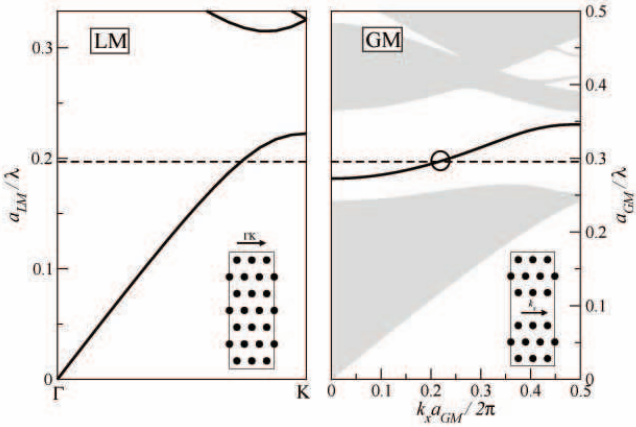

Fig. 4. (Left panel) Photonic band structure of the photonic crystal employed as lens material (LM) along the $\Gamma K$ direction (see inset). (Right panel) Projected band structure of TM modes in the guide material (GM). The gray regions represent the continuum of extended states in the photonic crystal. The white regions are the photonic band gaps. The wave-guide is formed by removing one row of Si rods as shown in the inset. The black line defines the guided modes inside the wave-guide. The frequencies in both panels are given in reduced units of the respective lattice parameters, $a_{LM}$ and $a_{GM}$. The horizontal dashed line defines the working frequency of optical devices under design.

The focusing effect can clearly be observed as represented in fig. 3. Here $S_x$ is plotted along the direction parallel to the $y$-axis for three diffrent cases; At the $x$-coordinate of the focal point for the two lenses, and for the incident beam at $x=0$. The estimated amplification in the focus equals $9.4dB$ for the 13 layers device and $9.1dB$ for the 9 layers, i.e. the 4 extra layers add an amplification of 0.3dB. A fit of $S_x$ $(x_f, y)$ to a Gaussian curve along the $y$-direction for the 13 layers lens gives a beam waist radius of $w_0 = 0.30\lambda_0$. This means that the lens has a spot-size conversion of $3.30\lambda_0 : 0.30\lambda_0 \sim 11 : 1$. Notable is that this conversion and amplification is achieved in a distance of $\sim 3.5\lambda_0$. This low F-number is difficult to achieve by conventional lens design. According to classical optics the F-number is defined by the focal distance divided by the width of the Gaussian beam, which here takes a value of only 0.47. This is achieved with a very low power loss, which is defined as the ratio of the transmitted power calculated inside the focus waist ($2w_0$) to that of the incident beam:

$$P_{loss} = 10 \log \left( \frac{\int_{-\omega_0}^{+\omega_0} [S_x(x_f, y)]_{lens} \, dy}{\int_{-\infty}^{+\infty} [S_x(0, y)]_{ext} \, dy} \right) \quad [dB] \quad (6)$$

The calculated value equals $1.0dB$, in other terms, 79% of the incident power is squeezed and passes through the focus.

### B. Photonic Crystal Wave-Guide Coupler

The lens demonstrates the possibility to stretch an incident Gaussian beam to a tenth of its width, sufficient for classical-PC wave-guide mode coupling requirements. In other words, if placing a single line-defect PC wave-guide at a distance of $4\lambda_0$ from the first file of cylinders in the device we should be able to couple that light to the wave-guide mode. This second inverse problem was set-up under very similar constrains as for the lens device. The lattice parameter of the PC, which we name guide-material (GM), is set so we have a guided mode for the $\lambda_0$ along the $\Gamma K$-direction, e.i. $a_{GM} = 1.5a_{LM}$. Right panel of fig. 4 shows how the working wavelength $\lambda_0$ of the lens correspond to a guided mode in the projected band structure for the infinite crystal.

If the wave-guide is placed at the focal point of the lens, one obtains a total insertion loss as high as $7.03dB$. This efficiency is calculated as the ratio of the total power that is transmitted through the wave-guide to the incident power (for coordinates references see fig. 5);

$$P_{loss} = 10 \log \left( \frac{\int_{-6a_{LM}}^{+6a_{LM}} [S_x(x = 35a_{LM}, y)]_{coupler} \, dy}{\int_{-\infty}^{+\infty} [S_x(0, y)]_{ext} \, dy} \right) \quad [dB] \quad (7)$$

In order to reduce the insertion loss we have solved the problem using two different ID set-ups. As a first approach, the inverse problem is constrained to a set of 52 lattice points, which are placed in front of the entrance. The idee of this mouth structure is to facilitate the coupling to the wave-guide mode in a taper-like manner. The lattice parameters are chosen as for the wave-guide. The fitness of the optimization was set equal to the sum of $S_x$ calculated at 30 points located along a transversal segment defined at the end of the wave-guide; i.e., on the segment $[-5.2a_{LM}, +5.2a_{LM}]$. Each individual has only 32 binary parameters, and just $2^{32} \sim 3 \times 10^9$ possible combinations are available. A resulting power loss of $1.07dB$ is obtained, which represents a substantial advance in comparison with the total loss when no mouth-structure was considered. This value equals 99% coupling efficiency with respect to the total energy that passes through the focus of the lens device. The resulting electric field amplitude and crystal structure is shown in the left panel in fig. 5.

In the second approach the optimization was carried out using the two crystal lattices simultaneously, the lattice used in the lens design plus the lattice used in the mouth design. This results in a total of $162 + 32$ binary parameters and, consequently, the space search size has been increased to $2^{162+32} \sim 2.5 \times 10^{58}$ possible solutions. To increase the probabilities of finding a global maximum in the optimization process we used an exaggerated population of 400 individuals.

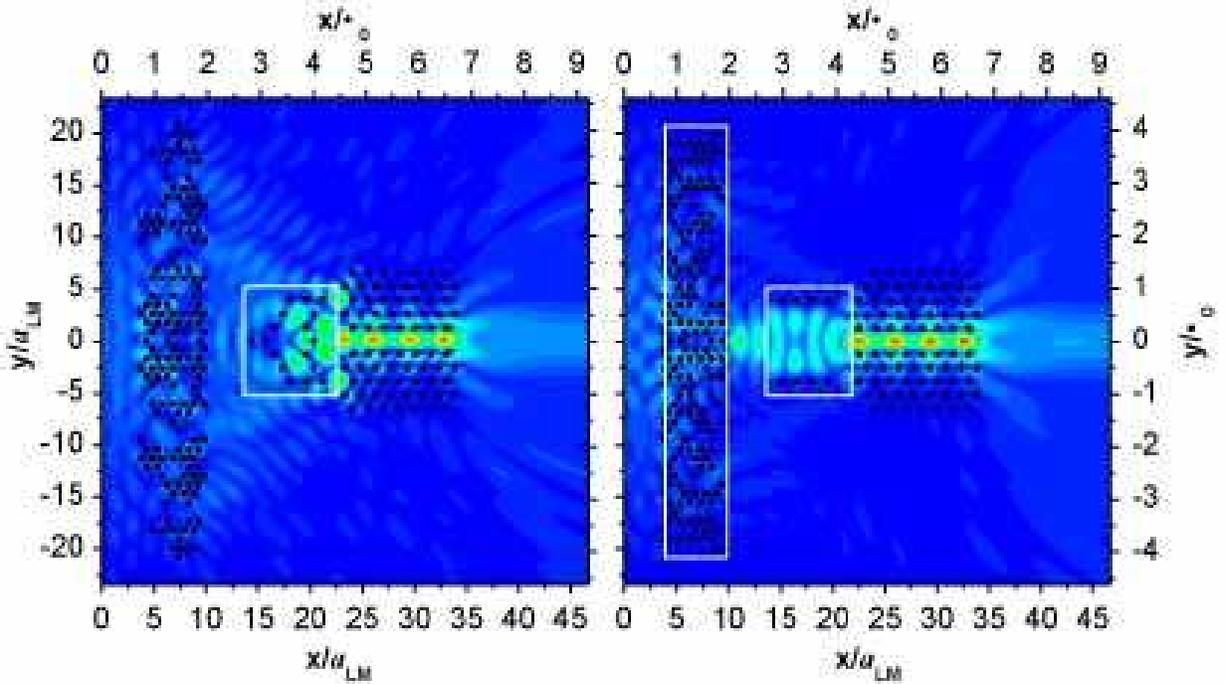

Fig. 5. Two optimized wave-guide coupler devices (dots) obtained by a genetic algorithm. The white rectangles enclose the cylinders representing the optimized cluster. (Left) The 13 layers lens from fig. 2 is used to narrow the incident beam. The marked cluster is optimized to facilitate the coupling to the wave-guide mode. A coupling loss of $1.07 dB$ is achieved. (Right) The crystal is a result of an optimization of the lens structure and the mouth cluster simultaneously to maximize the coupling efficiency. A coupling loss of $0.61 dB$ is achieved. A scale color is used; red (blue) color defines the maximum (minimum) electric field modulus.

This big population results in a slow convergence but, at the same time, an increase in the probabilities of fining the global optimum. For this design we needed 500'000 simulations to finish the run. This correspond to an effective calculation time of 30 days using a Pentium IV processor. The insertion loss predicted for this structure seen in the right panel in fig. 5, is as low as $0.61 dB$. This means that 87% of the impinging light passes through the wave-guide and is detected at the output. In fact, this value is underestimated since it does not include the light reflected at the end of the wave-guide by finite size effect. Therefore, the coupling efficiency predicted by this new structure is comparable with the better ones reported in the literature; Spuhler *et al.* [15] proposed an inverse designed wave-guide to fiber coupler with an efficiency improvement of 2dB per converter. References [16]–[18] report different tapered wave-guide couplers that exceed a predicted coupling efficiency of 90%. Finally, the J-coupler proposed by Prater *et al.* [19] predicted a coupling efficiency of 91%.

## IV. ROBUSTNESS OF THE DESIGN

An important issue is how this wave-guide coupler device will respond for a change in the frequency. Figure 6 plots the fitness for the frequencies of the guiding mode for the PC wave-guide. We see a sharp peak at the optimized frequency. This behavior is expected since the PC wave-guide coupler works due to multiple scattering between cylinders. Each cylinder in the cluster has an optimized position and contributes to the efficiency of the device. Therefore a large sensitivity to the working frequency of the device is expected. This functionality is much more sensitive than a tapered wave-guide [17]. Nevertheless, a specific PC wave-guide coupler working over a given range of frequencies can be designed by using this very same method.

A second issue of critical interest, due to the possible errors in fabrication, is to know the robustness of the device against fluctuations in the cylinders' radius or small displacements of their position in the lattice. Here we analyzed the robustness of the design photonic crystal based wave-guide coupler. Figure 7 shows the corresponding fitness value for 20 different calculated defected structures and for 5 different levels of relative error of their position (upper) and of the radius (lower). The structure is clearly more sensitive to defects due to displacement of the cylinders than to changes of their radiuses. This more rapid decrease in quality for errors associated with the position can be explained by a bigger absolute error, since the lattice constant is much larger than the radius of the cylinders. Notify that the fitness only is an estimation of the quality. The efficiency of each structure has to be calculated in more detail as explained earlier, integrating the pointing vector over a segment at the end of the wave-guide (see eq.

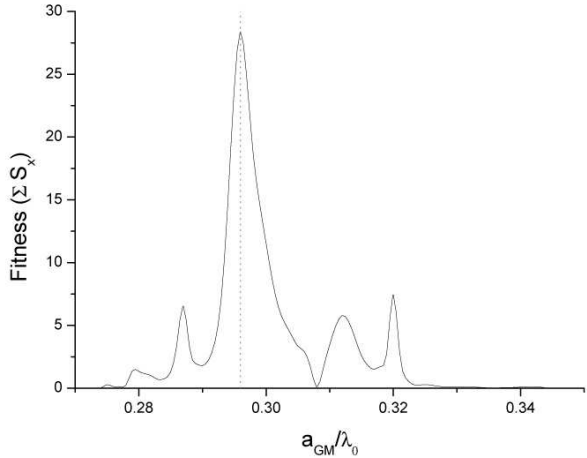

Fig. 6. The fitness of the optimized crystal wave-guide coupler is plotted for the frequencies in the gap of the PC wave-guide. The dotted line marks the frequency used in the optimization.

7). Calculating this value for the best crystal structure of the 20 simulated and for a relative error of 5%, we predict an efficiency of 84% and 65% for an error in the radius and lattice constant respectively. This sensibility, especially for displacements, is rather expected. As mentioned earlier, we are looking at a very complex and delicate system where each cylinder is given an optimized position and play an important role for the coupling effect. Now, this can of course be altered knowing the limitation of the manufacture procedure and introduce these flaws in the optimization procedure. Instead of optimizing a perfect system where the cylinders' positions and their radius' are fixed, we can easily solve the inverse problem and look for solutions which are more robust against the imperfection in the crystal, optimizing the expected structure from a fixed fabrication method. This will most likely lead to a decrease in the theoretical result but an increase in the experimental one.

Although our simulations involved the simplifying assumption of 2D PC's, it should be noticed that such 2D-periodic crystals can be studied in actual 3D crystals. For example, a recent simulation of 2D prism [8] reproduce fairly well the behavior of the actual structures in the micro wave regime. We conjecture that similar results might also be obtainable in the optical regime by using a PC-slab sandwiched between multilayer films with a large gap. Also, the preceding discussion focused on the TM modes of a structure based on "dielectric-scatterers-on-background". However, based on the general method presented here similar devices based on "holes-in-dielectric" structures can also be designed.

## V. Summary

In this work we have examined how inverse design of photonic crystals can be a very useful tool to optimize the performance of different PC devices and to find new unexpected properties. As a demonstration of the strength of the model a lens and a wave-guide coupler was design. Both structures show great new features and good quality in their performance. The robustness of the coupler device was also analyzed and discussed with respect to the inverse design setup. To conclude, this integration of MST with GA optimization can be used to solve a great diversity of inverse design problems in the field of PCs.

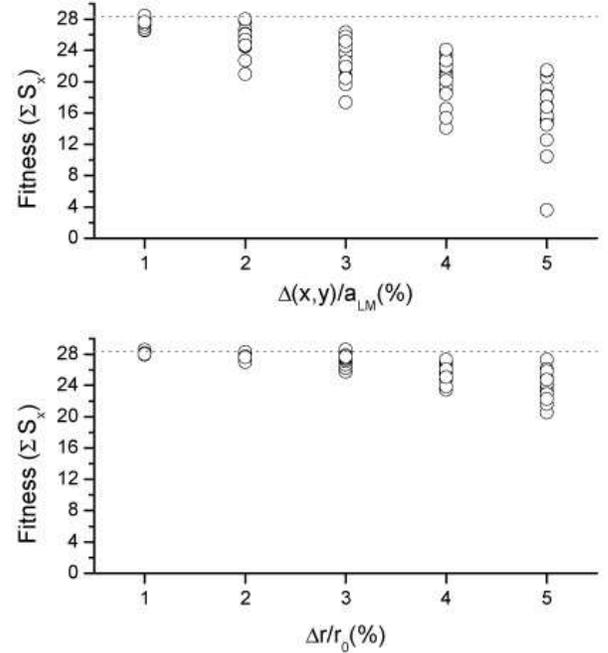

Fig. 7. Robustness of the design PC wave-guide coupler. The fitness for 20 structures for each level of error is plotted in the graph. The dotted line marks the fitness of the perfect structure. (Upper panel) Five different levels of relative error (1%-5%) in the position of the cylinder off their fixed lattice positions are shown. (Lower panel) Five levels of error in the radius of the cylinders are shown.

### Acknowledgment

The authors acknowledge financial support provided by the Spanish MED (Project number TEC2004-03545/MIC)

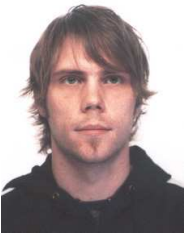

**Andreas Håkansson** was born in Göteborg, Sweden in 1976. He received the degree in physical and electrical engineering from the Technical University of Linköping, Sweden in 2002. He has been working toward a Ph.D. degree since 2002 at the Autonomous University of Madrid and at The Polytechnic University of Valencia.
Personal webpage: ttt.upv.es/andreas

**José Sánchez-Dehesa** was born in Toledo, Spain, in 1955. He received the Doctoral degree in Physics from the Autonomous University of Madrid, Spain in 1982. From 1985 to 2003 has been associated professor at the Autonomous University of Madrid. Now, he is full professor at The Polytechnic University of Valencia, and is the head of The Wave Phenomena Group associated to the Valencia Nanophotonic Technology Center. He has worked an semiconductors and low-dimensional structures such as quantum wells, super-lattices, quantum wires, etc. Now his work is mainly related with the study of photonic and phononic crystal devices.

**Lorenzo Sanchis** No biography availible.